\documentclass[prb,twocolumn,showpacs,superscriptaddress,groupedaddress]{revtex4}
\usepackage{epsfig,color}
\usepackage{dcolumn}
\newcommand{\p}{\mathrm{\pi}}
\newcommand{\beq}{\begin{equation}}
\newcommand{\eeq}{\end{equation}}
\newcommand{\bea}{\begin{eqnarray}}
\newcommand{\eea}{\end{eqnarray}}

\begin{document}
\title{Theory of Raman response in a superconductor with
           extended $s$-wave symmetry: Application to Fe-pnictides}

\author{Andrey V.~Chubukov}
 \affiliation{Department of Physics, University of Wisconsin-Madison, Madison, WI 53706, USA}
\author{Ilya Eremin}
 \affiliation{Max-Planck-Institut f\"{u}r Physik komplexer Systeme, D-01187 Dresden, Germany}
 \affiliation{Institute f\"{u}r Mathematische und Theoretische Physik, TU Braunschweig, D-38106 Braunschweig, Germany}
\author{Maxim M.~Korshunov}
 \affiliation{Max-Planck-Institut f\"{u}r Physik komplexer Systeme, D-01187 Dresden, Germany}
 \affiliation{L.V. Kirensky Institute of Physics, Siberian Branch of Russian Academy of Sciences, 660036 Krasnoyarsk, Russia}

\date{\today}

\begin{abstract}
We argue that Raman study of Fe-pnictides is a way to unambiguously
distinguish between various superconducting gaps proposed for these
materials. We show that $A_{1g}$ Raman intensity develops a true
resonance peak below $2\Delta$ if the pairing gap has $A_{1g}$
symmetry in the folded Brillouin zone ($\Delta (k=0) = \Delta,
~\Delta (\pi,\pi) = - \Delta$).  No such peak develops for a pure
$s$-wave gap, a $d$-wave gap, and an extended $s$-wave gap with
$\Delta(\mathbf{k}) = \Delta \cos{\frac{k_x}{2}}
\cos{\frac{k_y}{2}}$.  We show that the peak remains quite strong
 for the values of inter-pocket impurity scattering used to fit NMR data.
\end{abstract}


\maketitle

Recent discovery of superconductivity in the iron-based layered
pnictides with $T_c$ reaching 55K generated an enormous interest in
the physics of these materials~\cite{kamihara}. Most of
ferropnictides are quasi two-dimensional materials, and their parent
(undoped) compounds are metals and display antiferromagnetic
long-range order below $T_N \sim 150$K~\cite{kamihara,cruz,klauss}.
Superconductivity occurs upon doping of either electrons or holes
into FeAs layers, or by applying pressure. The electronic structure
measured by angle-resolved photoemission (ARPES)~\cite{kaminski} and
by magneto-oscillations~\cite{coldea} consists of two small hole
pockets centered around the $\Gamma=(0,0)$ point and two small
electron pockets centered around the $M=(\pi,\pi)$ point in the
folded Brillouin zone (BZ). The sizes of electron and hole pockets
are about equal in parent compounds.

The key unresolved issue for the pnictides is the symmetry of the
superconducting gap. A conventional phonon-mediated $s$-wave
superconductivity is unlikely because electron-phonon coupling
calculated from first principles is quite small~\cite{phonons}.
Several authors considered magnetically mediated pairing based
either on  itinerant~\cite{Mazin,others,li_new} or localized spin
models~\cite{bernevig} and argued that the gap should have an
extended $s$-wave symmetry $\cos k_x + \cos k_y$  (also called $s^+$
or, equivalently, $A_{1g}$ symmetry). This gap changes sign between
hole and electron pockets but has no nodes along the Fermi surface
(FS). On the other hand, another RPA study of magnetically mediated
superconductivity in the five-band Hubbard model~\cite{graser}
yielded two nearly degenerate candidate states in which the gap has
nodes on one of the FS sheets: either an extended $s$-wave state
with $\Delta(\mathbf{k}) \approx \Delta \cos{\frac{k_x}{2}}
\cos{\frac{k_y}{2}}$, or a $d_{x^2-y^2}$ state with
$\Delta(\mathbf{k}) \approx \Delta \sin{\frac{k_x}{2}}
\sin{\frac{k_y}{2}}$ (in the unfolded BZ, these two states are $\cos
q_x + \cos q_y$ and $\cos q_x - \cos q_y$,
respectively~\cite{comm}).

The experimental situation is also controversial.
 ARPES~\cite{kondo,ding} and Andreev spectroscopy~\cite{chen}
measurements have been interpreted as evidence for a nodeless gap,
either pure $s$-wave or $s^+$-wave. The resonance observed in
neutron measurements~\cite{osborn} is consistent with the $s^+$
gap~\cite{Korshunov}. On the other hand, nuclear magnetic
resonance (NMR) data ~\cite{Nakai} and some of
penetration depth data~\cite{pen_depth} were interpreted
as evidence for the nodes in the gap. Some, but not all of the data
can be reasonably fitted by a model of an $s^+$ SC with ordinary
impurities~\cite{mazin_imp,Chubukov,anton}.

In view of both theoretical and experimental uncertainties, it is
important to find measurements which could unambiguously
distinguish between different pairing symmetries. Recent suggestions
for such probes include Andreev bound state~\cite{ashvin} and
Josephson interferometry~\cite{mazin_parker}. In this communication,
we argue that the study of $A_{1g}$ Raman intensity is another way
to determine the symmetry of the superconducting gap. We show that
 in the $A_{1g}$ scattering geometry
 the  Raman signal develops a true resonance below $2\Delta$ for
the case of $s^+$ gap. No such resonance appears for a pure $s$-wave
gap, for $\cos{\frac{k_x}{2}} \cos{\frac{k_y}{2}}$ and
$\sin{\frac{k_x}{2}} \sin{\frac{k_y}{2}}$ gaps.
 The $A_{1g}$ resonance is the effect of the final state interaction,
which is known to be important for Raman
scattering~\cite{final_state}. A similar resonance occurs in the
$B_{1g}$ channel in a magnetically mediated $d_{x^2-y^2}$
superconductor~\cite{resonance_raman}, but there the resonance is
weakened by a finite damping associated with nodes of the $d$-wave
gap.

We model Fe-pnictides by an itinerant electron system with two
(almost) degenerate hole FS pockets centered at the $\Gamma$ point and
two electron FS pockets centered at the $M$ point. We assume that the
 magnitude of the gap $\Delta$ is much smaller than the Fermi
energy. In this situation, Raman intensity
at frequencies $\leq 2\Delta$ is determined by states near the FS
where the density of states (DOS) can be approximated by a constant.
 We first assume that the pairing gap has $s^+$ symmetry,
$\Delta(\mathbf{k} \approx 0) = \Delta$, $\Delta(\mathbf{k} \approx
\p) = -\Delta$, and show how the resonance appears. We then discuss
other pairing symmetries.

Without final state interaction, the Raman intensity in a clean BCS
$s^+$ superconductor is the same as in a pure $s$-wave
superconductor\cite{final_state} and is given by $I_i(\Omega) =
2\mathrm{Im}R_i(\Omega)$, where
\begin{eqnarray}
R_{A_{1g}}(\Omega) = -R_0 &&\left< \int d\omega \gamma^2_{A_{1g}}
\left[1-\frac{\omega_+\omega_- -\Delta^2}{\sqrt{{\omega}^2_+ -\Delta^2}
\sqrt{{\omega}^2_- -\Delta^2}}\right] \right. \nonumber\\
&& \times \left. \frac{1}{\sqrt{{\omega}^2_+
-\Delta^2}+\sqrt{{\omega}^2_- -\Delta^2}} \right>_{FS}
\label{r_7}
\end{eqnarray}
Here $\gamma_{A_{1g}} = \cos k_x + \cos k_y$ is the geometrical
factor for $A_{1g}$ scattering, $\omega_\pm = \omega \pm \Omega/2$,
and $\left<...\right>_{FS}$ denotes the averaging over FS.  The
factor $2$ in the relation between $I_i(\Omega)$ and $R_i(\Omega)$
reflects the fact that there are two hole and two electron pockets.
Other factors are incorporated into $R_0$.

\begin{figure}[tbp]
\includegraphics[angle=0,width=1.0\linewidth]{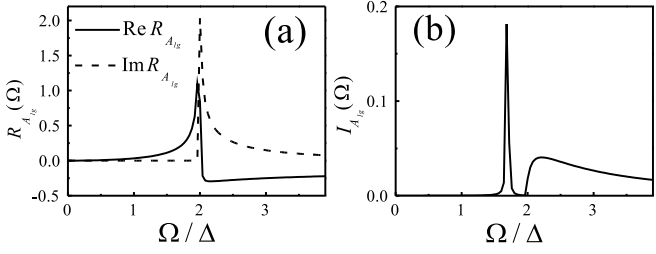}
\caption{Real and imaginary parts of the $A_{1g}$ Raman intensity
for a clean $s^+$ superconductor without (a) and with (b) final
state interaction. Final state interaction gives rise to a
well-defined resonance in the $A_{1g}$ intensity. We used $R_0 =
1/(4\pi)$, $u_{eff} R_0 \approx 0.4$, and added damping $\gamma =
0.001\Delta$.} \label{fig1}
\end{figure}

\begin{figure}[t]
\includegraphics[angle=0,width=1.0\linewidth]{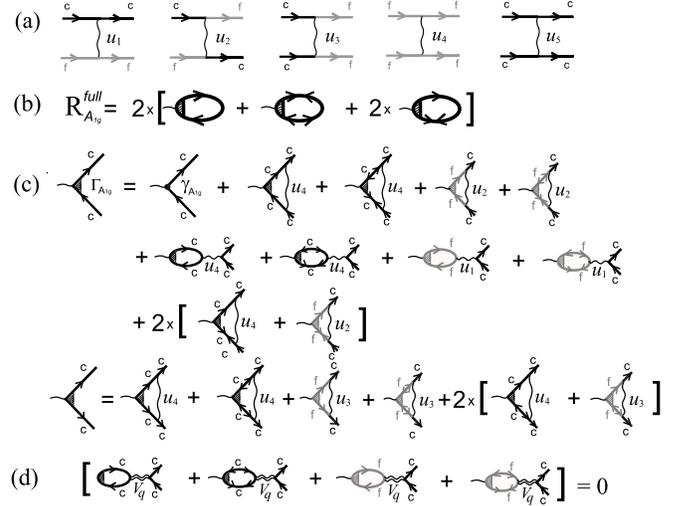}
\caption{(a) Five relevant interactions between fermions near hole
and electron FS pockets. Black and grey lines represent fermionic
$c$- states near $(0,0)$ and $f$- states near $M=(\pi,\pi)$. (b) The
full Raman bubble, which is the sum of $GG$, $FF$ and $GF$ terms.
Only the contribution from $c-$fermions is shown. The one from
$f$-fermions is obtained by replacing $c$ lines by $f$-lines and
vice versa in (b) and (c) panel. (c) The renormalization of the
$A_{1g}$ Raman vertex for $c$- fermions. The first 8 diagrams
account for ``conventional'' renormalization of the $A_{1g}$
particle-hole vertex and involve $GG$ and $FF$ bubbles, the last two
diagrams involve $GF$ bubbles and emerge due to a non-zero coupling
between $A_{1g}$ particle-hole channel and ordinary $s-$wave pairing
channel.  The renormalization of the particle-particle vertex in
turn involves 4 ``conventional'' diagrams with $GG$ and $FF$ terms,
which account for the renormalization in the particle-particle
channel, and two diagrams due to the coupling to the $A_{1g}$
particle-hole channel. (d) The renormalizations due to long-range
component of Coulomb interaction $V_{q} \propto 1/q^2$. This
renormalization vanishes because of the symmetry between $c-$ and
$f-$fermions and the fact that $\gamma_{A_{1g}} (k=0) = -
\gamma_{A_{1g}} (\pi)$. } \label{fig2}
\end{figure}

The intensity $I_{A_{1g}}(\Omega)$ computed using (\ref{r_7})
vanishes at $\Omega < 2\Delta$ and is discontinuous at $2\Delta$.
The real part of $R_{A_{1g}}$, which we will need later, is positive
below $2\Delta$, scales as $\Omega^2$ at small frequencies, and
diverges upon approaching $2\Delta$ from below~\cite{ch_n}. We show
both $\mathrm{Re}R_{A_{1g}}$ and $\mathrm{Im}R_{A_{1g}}$ in
Fig.~\ref{fig1}.
The final state interaction is diagrammatically represented as the
renormalization of the Raman vertex. Vertex corrections arise from
multiple insertions of fermion-fermion interactions into the Raman
bubble. There are five different interactions between low-energy
fermions (see Fig.~\ref{fig2}(a)). They include intra-band
interactions for electrons ($u_4$) and for holes ($u_5$), which we
assume to be equal, inter-band interactions $u_1$ and $u_2$ with
momentum transfer $0$ and $(\pi,\pi)$, respectively, and the pair
hopping term $u_3$.

A generic theory of vertex renormalizations has been developed in
Ref.\cite{final_state} and we follow this work in our analysis.  In
general, there are three different types of vertex corrections: (i)
the corrections which come from short-range interactions $u_i$ and
transform a bare $A_{1g}$ Raman vertex into a renormalized {\it
particle-hole} vertex (these corrections involve $GG$ and $FF$
bubbles, where $G$ and $F$ are normal and anomalous Green's
functions), (ii) the corrections which transform a Raman vertex into
a particle-particle vertex (these involve $GF$ bubbles) (see Fig.
\ref{fig2}(b)), and (iii) the corrections from the long-range
component of the Coulomb interaction $V_{q} \propto 1/q^2$.  The
corrections of the first type are given by ladder and bubble
diagrams which involve $u_1$, $u_2$ and $u_4$ vertices (first 8
terms in Fig.~\ref{fig2}(c)). The corrections of the second-type are
non-zero when the  symmetry of the gap is the same as the symmetry
of the Raman vertex, which is our case.  These corrections transform
$A_{1g}$ particle-hole vertex into an ordinary $s-$wave pairing
vertex (the terms with the overall factor 2 in Fig.~\ref{fig2} (c)).
The third renormalization, due to long-range component of the
Coulomb interaction, generally  gives rise to a screening of the
Raman signal~\cite{final_state,abr}, but vanishes in our case
because of particle-hole symmetry and the fact that $A_{1g}$ Raman
vertex $\gamma_{A_{1g}}$ changes sign between hole and electron
pockets (see Fig. \ref{fig2}(d)).
Note in this regard that $V_{q} \propto 1/q^2$ is not a part of RG
transformation and depends only on a momentum transfer $q$, in
distinction to the other two interactions with small momentum
transfer, $u_4$ and $u_1$. The bare values of $u_4$ and $u_1$  may
be identical, but the two flow in different directions under RG.
Also note that we did not include a momentum-independent term into
$\gamma_{A_{1g}}$. If $\gamma_{A_{1g}}$ had such component, it would
be screened by the long-range Coulomb interaction.

Combining the renormalizations (i) and (ii) and evaluating the
diagrams, we obtain the full Raman intensity
$I^{full}_{A_{1g}}(\Omega) = 2\mathrm{Im}R^{full}_{A_{1g}}(\Omega)$
with $R^{full}_{A_{1g}} (\Omega)$ in the form
\begin{eqnarray}
\lefteqn{R^{full}_{A_{1g}} (\Omega) =}& &  \nonumber \\
&& \frac{R_{A_{1g}} (\Omega) \left( 1 + u_f R_{pp} (\Omega)\right)+
4 u_f R^2_{mix} (\Omega)}{(1 - u_{eff} R_{A_{1g}} (\Omega))(1 + u_f
R_{pp}(\Omega)) - 4 u_g u_f R^2_{mix} (\Omega)}, \nonumber
\\
\label{n_1}
\end{eqnarray}
where $u_{eff} =  2 u_1 - u_2 - u_4$ is the effective vertex for the
Raman renormalization in the $A_{1g}$ particle-hole channel,
$u_{f}=u_3+u_4$, $u_g=u_4-u_2$,  $R_{pp}(\Omega) \propto
\log{E_F/\Omega} >0$ is the polarization bubble in the $s-$wave
particle-particle channel, and $R_{mix} (\Omega) \propto \int d^2 k
d \omega \gamma_{A_{1g}} G_{k, \omega + \Omega} F_{k, \omega}$
couples $A_{1g}$ particle-hole channel and $s$-wave
particle-particle channel. At low frequencies, $R_{mix} (\Omega)
\propto \Omega$.  In this respect, the situation is similar to the
case of a spin resonance in a $d-$wave superconductor, where $S=1$
particle-hole channel couples to $S=0$ particle-particle
channel~\cite{coupl}.

Because $s$-wave channel is repulsive in our case ($u_3 + u_4 >0$),
there is no pole in $R^{full}_{A_{1g}} (\Omega)$ coming from the
particle-particle channel. Furthermore, $ R_{pp} (\Omega)$
logarithmically diverges at $\Omega << E_F$, and canceling this
divergent term between the numerator and the denominator in
(\ref{n_1}), we obtain
\begin{eqnarray}
I^{full}_{A_{1g}} (\Omega) \approx  2 \frac{\mathrm{Im}R_{A_{1g}}}
{\left(1-u_{eff} \mathrm{Re}R_{A_{1g}}\right)^2 + \left(u_{eff} \mathrm{Im}R_{A_{1g}} \right)^2}.
\label{4}
\end{eqnarray}
We see therefore that the coupling between $A_{1g}$ particle-hole
and $s-$wave particle-particle channels is irrelevant, and the full
$I^{full}_{A_{1g}} (\Omega)$ can be  approximated by the expression
which only includes vertex corrections which preserve particle-hole
structure of the Raman vertex.

Our next observation is that for two-band structure, $u_{eff}$
contains the terms $u_1$ and $u_2$, which do not contribute to the
renormalization of the $s^+$ pairing vertex (the latter involves
$u_3$ and $u_4$ terms~\cite{Chubukov}), i.e., in distinction to
one-band case~\cite{final_state}, the renormalization of the
$A_{1g}$ Raman vertex and the renormalization of the $s^+$ pairing
vertex (which has the same $A_{1g}$ symmetry)
 are given by different combinations of the interactions $u_i$

Finally, we note that  below $2\Delta$, $\mathrm{Im}R_{A_{1g}}=0$ while
$\mathrm{Re}R_{A_{1g}}$ is positive and evolves between zero and
infinity when $\Omega$ changes between zero and $2\Delta$. Then, for
 positive $u_{eff}$,  the $A_{1g}$ Raman intensity develops a
$\delta$-functional resonance peak below $2\Delta$, at a frequency
where $u_{eff} Re R_{A_{1g}} =1$. For a $d_{x^2-y^2}$ superconductor
the same effect leads to an excitonic resonance in a staggered spin
susceptibility~\cite{resonance}, and to a pseudo-resonance in a
$B_{1g}$ Raman response~\cite{resonance_raman}.

The flow of the interactions between the bandwidth $W$ and the Fermi
energy $E_F$ has been analyzed in the earlier RG
study~\cite{Chubukov}, and the result is that $u_1$ becomes the
largest interaction at energies comparable to the Fermi energy, even
if the intra-band Hubbard repulsion $u_4$ is the largest term in the
Hamiltonian.  Specifically,  in the RG flow $u_1$ and $u_3$
increase, $u_2/u_1$ flows to zero, and the $u_4$ term  decreases,
such that $u_{eff}=2u_1-u_4-u_2$ becomes positive  at energies below
$E_F$, relevant to Raman scattering, and \textit{the $A_{1g}$ Raman
response develops a resonance below $2\Delta$}. We emphasize that
the physics which makes $u_{eff}$ positive is the same physics that
gives rise to an attraction in an extended $s^+$-wave pairing
channel. Indeed, the pairing interaction in $s^+$ channel, which is
the combination $u_3-u_4$, becomes positive under RG.

For other proposed gap symmetries, the resonance does not develop,
even if one neglects the screening by long-range Coulomb
interaction. For an $s$-wave gap, there is no sign change between
electron and hole FS, and the analog of $u_{eff}$ in Eq.~(\ref{4})
is $-2u_1-u_4+u_2$. This combination is negative, so the resonance
does not occur. For a gap that changes sign along either hole or
electron FS, the largest contribution to $I_{A_{1g}}(\Omega)$ comes
from the FS along which the gap is nodeless. Vertex renormalization
for such term contains $-u_4+(2u_1-u_2)x$, where $x \sim k_F^2$, and
$k_F$ is a small radius of the FS along which the gap has nodes. In
this situation, $u_1$ term does not overcome $u_4$, and the
resonance does not occur. For $d_{xy}$ gap ~\cite{li_new} with
$\Delta \propto \sin k_x \sin k_y$ (in the folded BZ), all $u_i$
terms in the vertex renormalization are reduced. Resonance may still
occur, but the effective interaction now is small, $O(k^2_F)$, and
the resonance is washed out by a small damping. This shows that the
$A_{1g}$ Raman resonance is a fingerprint of an $s^+$ pairing.

Finally, we consider how the resonance in $s^+$ superconductor is
affected by ordinary impurities. As in earlier
works\cite{Chubukov,anton}, we introduce impurity potential $U_i(q)$
with intra- and inter-pocket terms $U_i(0)$ and $U_i(\p)$,
respectively and restrict with the Born approximation. This
approximation (which requires $U_i << E_F$) may not work for $U_i
(0)$ (Ref.~\cite{mazin_imp}) but should be  valid  for $U_i (\pi)$
which is pair-breaking and is therefore very likely not larger than
$\Delta << E_F$. For our case, $U_i (0)$ controls the functional
form of  $Re R_{A_{1g}}$, which still evolves between zero and
infinity when $\Omega$ changes between $0$ and $2\Delta$,  while the
broadening of the resonance is entirely due to $U_i (\pi)$.  In this
situation, Born approximation should be sufficient.

The calculations are straightforward and we refrain from presenting
the details. Intra-pocket impurity scattering does not affect the
gap by Anderson's theorem, but $U_i(\p)$, which scatter fermions
with $+\Delta$ and $-\Delta$, is pair-breaking and affects the gap
in the same way as magnetic impurities in an ordinary $s$-wave
superconductor. We use $b = 2U_i(\p)/\Delta$, where $\Delta$ is the
order parameter as a measure of the strength of pair-breaking
impurity scattering.
\begin{figure}[tbp]
\includegraphics[angle=0,width=1.0\linewidth]{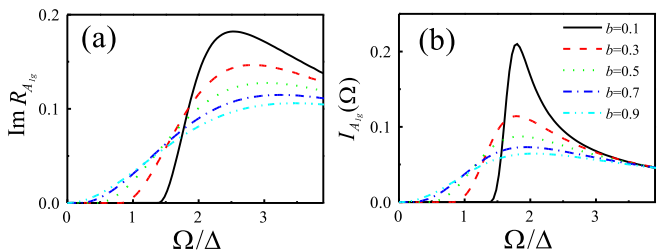}
\caption{(color online) Calculated Raman intensity for an $s^+$
superconductor without (a) and with (b) final state interaction for
various strength of the inter-band impurity scattering. We use the
same $u_{eff}$ as in Fig. \protect\ref{fig1} and for definiteness
set $U_i(0)=\Delta$.} \label{fig3}
\end{figure}

The results of the calculations are shown in Fig.~\ref{fig3}, where
we plot Raman intensity in the presence of impurities both without
and with final state interaction. Comparing this
figure with Fig.~\ref{fig1} we see that the resonance gets damped at
a finite $b$, and Raman intensity no longer shows two peaks. Still,
the resonance continue to determine the shape of
$I_{A_{1g}}(\Omega)$: without final state interaction the peak
broadens and shifts to larger frequencies $\Omega > 2\Delta$ upon
increasing $b$, while when the final state interaction is included,
the peak remains below $2\Delta$ and shifts to a smaller frequency
with increasing $b$.  Notice that the resonance is still
quite strong at $b \sim 0.5-0.7$, which was used to fit NMR and
penetration depth data~\cite{Chubukov, anton}. In other words, it
should be observable in Raman experiments if indeed the gap has an $s^+$
symmetry.

To conclude, in this paper we argued that Raman study of
Fe-pnictides is a way to unambiguously distinguish between various
superconducting gaps proposed for these materials. We have shown
 that for an $A_{1g}$ ($s^+$) gap $\Delta(\mathbf{k} \approx
0) = \Delta$, $\Delta(\mathbf{k} \approx \p) \approx - \Delta$, the
$A_{1g}$ Raman intensity has a true resonance peak below $2\Delta$.
No such peak emerges for a pure $s$-wave gap, a $d_{x^2-y^2}$ gap,
and an extended $s$-wave gap with $\Delta(\mathbf{k}) = \Delta
\cos{\frac{k_x}{2}} \cos{\frac{k_y}{2}}$. The resonance peak gets
broader by pair-breaking inter-pocket impurity scattering but is
still fairly visible for the values of impurity scattering used to
fit NMR data.

We acknowledge useful conversations with G. Blumberg, W. Brenig,
H.-Y. Choi, D.V. Efremov, A. Sacuto, M. Vavilov, A. Vorontsov.
A.V.C. acknowledges support from NSF-DMR 0604406. I.E. acknowledges
partial support from the Asian-Pacific Center for Theoretical
Physics, the Volkswagen Foundation (I/82203)
 and the Program "Development of scientific potential of a higher school"
(N1 2.1.1/2985). M.M.K. acknowledges support from RFBR 07-02-00226,
OFN RAS program on ``Strong electronic correlations'', and RAS
program on ``Low temperature quantum phenomena''.


\begin{thebibliography}{99}
\bibitem{comm} Strictly speaking, the extended $s$-wave states with
    $\Delta(\mathbf{k}) \propto \cos k_x + \cos k_y$ and $\Delta(\mathbf{k}) \propto
    \cos{\frac{k_x}{2}} \cos{\frac{k_y}{2}}$ cannot be decoupled in the
    gap equation as both are members of the $A_{1g}$ representation
    of $D_{4h}$ tetragonal space group. However, for small hole and
    electron pockets, the matrix element between the two is small.
\end{thebibliography}
\end{document}